\documentstyle[12pt,epsfig]{article}
\begin{document}
{\hskip 12.0cm} KIAS-P98023\par
{\hskip 12.0cm} SNUTP 97-151\par
\vspace{1ex}
\vspace{2ex}
\begin{center}        
{\LARGE \bf MeV Tau Neutrino in Gauge Mediated Supersymmetry Breaking Model}\\
\vspace{5ex}
{\sc Chun Liu}\footnote{\it email: liuc@kias.re.kr}\\
{\it Korea Institute for Advanced Study}\\
{\it 207-43 Chungryangri-dong, Dongdaemun-ku, Seoul 130-012, Korea}\\ 
\vspace{3ex}
{\sc H.S. Song}\footnote{\it email: hssong@physs.snu.ac.kr}\\
{\it Center For Theoretical Physics, Seoul National University}\\
{\it Seoul, 151-742, Korea}\\

\vspace{6.0ex}
{\large \bf Abstract}\\
\vspace{4ex}
\begin{minipage}{130mm}
                                                                               
   A supersymmetric model which naturally accommodates MeV tau neutrino within
the framework of gauge mediated supersymmetry breaking is described.  The 
lepton number violation is originally introduced in the messenger sector of 
the theory.  A large slepton-Higgs mixing mass and a small lepton-higgsino 
mixing mass are generated at one-loop.  Scalar tau neutrino has non-vanishing 
vacuum expectation value.  These results in a non-zero $\nu_{\tau}$ mass which 
is in the range of ($1-10$) MeV.\par  
\vspace{1.0cm}
\end{minipage}
\end{center}

\newpage

{\bf I. Introduction}\\

   Massive $\tau$-neutrino with mass in the range of $1-10$ MeV
is an interesting scenario for astrophysics and cosmology [1].
It should have lifetime of $0.1-100$ sec [1] or sufficient annihilation rate
[2].
As summarized in Ref. [1], it can relax the big-bang nucleosynthesis
bound to the baryon density and the number of neutrino species; allow big-bang
nucleosynthesis to accommodate a low ($< 20\%$) $^4$He mass fraction or high
($> 10^{-4}$) deuterium abundance; improve significantly the agreement between 
the cold dark matter theory of structure formation and observation; and help 
to explain how type-II supernova explodes.
\par
\vspace{1.0cm}    
   Due to the useful phenomenological consequence, it is interesting to see if
there is a natural way to accommodate the MeV tau neutrino.  It can be 
achieved by introducing right-handed neutrinos into the Standard Model (SM).  
To make the mass range natural, the Majorana mass scale of the right-handed 
neutrinos should be properly chosen.  And fermion family symmetry may be 
further introduced to keep the e-neutrino and $\mu$-neutrino being light. This
$\tau$-neutrino must decay
or annihilate,
{\it e.g.} into light neutrinos and massless boson
[3], fast enough to avoid the overclosure of the universe.  Nevertheless, the 
above logic essentially puts the explanation of the neutrino masses into the 
same category as that of the other fermion masses.
\par
\vspace{1.0cm}
   Within supersymmetry, which is the most favorable framework for physics 
beyond SM, neutrino masses can have several alternative origins.  This is 
simply because that in this case, the lepton number is no longer automatically
conserved at tree level.  By assuming the conservation of the baryon number
only, practically viable models can be constructed without contradiction to 
the current experiments.  They are the so-called R-parity violating models 
(with baryon number conservation).  In such models, the possible new origins 
of the neutrino masses can be typically classified into following scenarios.
First is the non-vanishing sneutrino vacuum expectation values (vevs) [4, 5].  
If the sneutrino vevs are non-zero, neutrino in general gains mass due to its
tree-level mixing with neutralinos.  A neutrino with the heaviness of several
MeV can be generated.
The second scenario is due to lepton number violating interactions in
superpotential.  There are two kinds of these interactions that are 
renormalizable, the bilinear and the trilinear terms.  In case the bilinear 
terms can be rotated away by redefining Higgs superfield [6], the lepton 
number violation can all be realized in the trilinear terms.  These trilinear 
interactions induce neutrino masses at the loop level [7].  However, the 
trilinear coupling constants are so constrained by some phenomenological 
considerations [8] that this mechanism cannot produce $\tau$-neutrino mass
larger than 1 MeV within reasonable range of supersymmetric mass scale.  
The third possibility lies in the soft supersymmetry breaking terms with 
lepton number violation [9].  The effect of lepton number violation will be 
mediated to neutrino masses through loops.  The simplest case is just to
introduce bilinear mass terms which mix the Higgs boson with scalar neutrinos.
They induce the mixing between neutrinos and higgsino, which in turn generates 
neutrino masses by see-saw mechanism.  For the soft masses being around weak
scale, neutrino mass of several MeV can be generated [6] 
\footnote{A recent discussion was made in Ref. [10].}.
Both the first and the third kinds of origin for MeV neutrino rely on the
deeper structure of the theory, namely the supersymmetry breaking mechanism,
because they are closely related to the soft breaking sector.
\par
\vspace{1.0cm}
   Within the framework of minimal supergravity, Ref. [11] studied the
R-parity violation characterized by bilinear terms in the superpotential.
They are the most relevant terms to heavy neutrino mass.  Generally, they
result in the sneutrino vevs which might be around $100$ GeV.  Such large
values, however, do not mean $100$ GeV heavy neutrino masses, because there
is an almost alignment in the mass matrix [5].  In other words, by a suitable
choice of basis, the bilinear terms are rotated away and the corresponding
soft terms are almost rotated away.  Effectively, there are only small
sneutrino vevs in this basis which can give $\tau$-neutrino mass ranging from
sub-eV to MeV.
\par
\vspace{1.0cm}
   In this paper, we consider the MeV $\tau$-neutrino as well as the R-parity
violation in the framework of gauge mediated supersymmetry breaking (GMSB).
We notice that it can be natural that even in the basis where the bilinear
terms in superpotential are absent, the theory still allows a relatively
large sneutrino vev which is about $1-10$ GeV.
For such a sneutrino vev, the lepton
number breaking must not be spontaneous.  Otherwise the corresponding 
Goldstone boson would result in unacceptable consequences both in astrophysics
[12] and in Z decays [13].  Some explicit lepton number violations have to be 
introduced further, like the soft supersymmetry breaking terms with lepton 
number violation.
This scenario generates MeV neutrino provided that the Zino mass is around
$100-1000$ GeV.  It will be realized in GMSB in next section.
\par
\vspace{1.0cm}
{\bf II.  The Model}\\

   In this section, we construct a simple model which accommodates MeV 
$\tau$-neutrino within the framework of GMSB.
The lepton number violation is introduced originally in the messenger
sector of the theory.  It then is communicated to the SM sector including the 
related soft supersymmetry breaking terms.  The $\tau$-neutrino mass appears
naturally in a way which combines the first and third scenarios described
above.  The explanation for the soft breaking mass terms with lepton number
violation is given.  
\par
\vspace{1.0cm}
   GMSB theory [14, 15] has drawn a lot of attentions recently.  Supersymmetry 
breaking is communicated from the hidden sector to the observable sector of 
the theory via gauge interactions.  The scale of supersymmetry breaking is 
comparatively low, so that the flavor changing neutral current processes are 
sufficiently suppressed.  When considering MeV $\tau$-neutrino, we will make 
use of the observations of Dine and Nelson [15], and Dvali {\it et al} [16].
They noted that in GMSB the $\mu$ problem [17] is rather severe.  Both the 
$\mu$ term which is the mixing mass term of the two Higgs doublets and its 
corresponding soft breaking $B_{\mu}$ term can be generated at one-loop [16].  
Either $\mu$ is at the weak scale and $B_{\mu}$ is unnaturally large, or 
$B_{\mu}$ is at the weak scale and $\mu$ is very small.  While there are 
possible solutions of this problem [18], we will not touch this problem in 
this work.  Instead, we apply similar observation to the discussion of 
another mixing term, that is the one between the lepton and the Higgs doublet.
\par
\vspace{1.0cm}
   We extend the model of Dine and Fischler [14] to include the lepton number 
violation.  To keep the other two neutrinos being light, a discrete family 
symmetry, which is a $Z_3$ symmetry among the SU(2) doublets of the three 
generations, is assumed.  The gauge group of the model is just 
SU(3)$\times$SU(2)$\times$U(1).  The supersymmetric gauge interactions are 
uniquely determined and can be found in text books.  Besides the fields of the 
particles in the minimum supersymmetric SM, like the left-chiral lepton 
superfields and their SU(3)$\times$SU(2)$\times$U(1) quantum numbers
$L_i(1, 2, -1)$ where $i=1, 2, 3$ for three families, the Higgs superfields
$H_u(1, 2, 1)$ and $H_d(1, 2, -1)$, additional set of chiral superfields, 
which is usually called messenger sector, is introduced:
\begin{equation}
S, S' = (1, 2, -1)~, ~~~~~\bar{S}, \bar{S'} = (1, 2, 1)~,
\end{equation}
and
\begin{equation}
T, T' = (3, 1, -2/3)~, ~~~~~\bar{T}, \bar{T'} = (\bar{3}, 1, 2/3)~.
\end{equation}
Furthermore, there are three gauge-singlet superfields, $X$, $Y$, and $V$.  
$Y$ is responsible for supersymmetry breaking, $X$ is related to electro-weak 
symmetry breaking, and $V$ to lepton number violation. 
\par
\vspace{1.0cm}
   The superpotential of the model is written as follows,
\begin{equation}
{\cal W}={\cal W}_1 + {\cal W}_2~,
\end{equation}
where ${\cal W}_1$ conserves lepton number,
\begin{equation}
\begin{array}{lll}
{\cal W}_1&=&m_1(\bar{S'}S+S'\bar{S})+m_2(\bar{T'}T+T'\bar{T})+m_3S\bar{S}
+m_4T\bar{T}+m_5V^2\\[3mm]
&&+Y(\lambda_1S\bar{S}+\lambda_2T\bar{T}+\lambda_3V^2-\mu_1^2)
+\lambda_4X(H_uH_d-\mu_2^2)~.
\end{array}
\end{equation}
In above equation, the Yukawa interactions are omitted which are irrelevant to 
our discussion.  ${\cal W}_2$ violates the lepton number but has $Z_3$ family
symmetry, 
\begin{equation}
{\cal W}_2 = V(\lambda_5H_uS+\lambda_6\sum_iL_i\bar{S})~.
\end{equation}
   The supersymmetry breaking is communicated to the observable sector by the 
messengers.  The physics related to ${\cal W}_1$ has been discussed thoroughly
in Ref. [14].  The only thing different is that we have introduced one more
gauge-singlet $V$.  For $m_5^2$ sufficiently large, $V$ does not develop any 
vev.  The form of the superpotential is not the most general one which 
follows the symmetry principle.  However it is natural in the sense of 
t'Hooft due to the non-renormalization theorem in supersymmetry.
$\mu_1$ is the supersymmetry breaking scale.  $\mu_2$ fixes the electroweak
scale, namely the vevs of Higgs fields.  It therefore contributes to higgsino
masses as will be seen explictly later.
\par
\vspace{1.0cm}
   For the superpotential ${\cal W}_2$, as can be seen, it is the second term 
of Eq. (5) that violates lepton number.  We have freedom to redefine 
$\frac{1}{\sqrt{3}}\sum_iL_i\equiv L_{\tau'}$ which can be regarded as the 
weak eigenstate of $(\nu_{\tau}, \tau)$ superfield, then
\begin{equation}
{\cal W}_2 = V(\lambda_5H_uS+\sqrt{3}\lambda_6L_{\tau'}\bar{S})~.
\end{equation}
It results in effective $\tau$ lepton number violating interactions by 
integrating out the heavy messengers, 
\begin{equation}
{\cal L}_{\rm eff}^{\not L} = \sqrt{3}\mu_{\tau}L_{\tau'}H_u|_{\theta\theta}
+\sqrt{3}B_{\mu_{\tau}}A_{\tau'}\phi_u +{\rm h.c.}~,
\end{equation}
where $A_{\tau'}$ and $\phi_u$ denote the scalar fields of the superfields
$L_{\tau'}$ and $H_u$, respectively, and both $\mu_{\tau}$ and 
$B_{\mu_{\tau}}$ are generated through one-loop given in Fig. 1,
\begin{equation}
\begin{array}{lll}
\mu_{\tau}&\simeq&\displaystyle
\frac{\lambda_5\lambda_6}{16\pi^2}\frac{\mu_1^2}{m_3}~,\\[3mm]
B_{\mu_{\tau}}&\simeq&\displaystyle\frac{\lambda_5\lambda_6}{16\pi^2}
\left(\frac{\mu_1^2}{m_3}\right)^2~.\\[3mm]
\end{array}
\end{equation}
It is easy to see from Eq. (4) that $\mu_1^2$ is the vev of the auxiliary
component of $Y$.  From Eq. (8), we have the relation,
\begin{equation}
B_{\mu_{\tau}} = \mu_{\tau}\frac{\mu_1^2}{m_3}~.
\end{equation}
$\mu_1^2$ is constrained by the soft masses of the superpartners of the 
particles in SM.  It is natural to take the messenger mass scale $10^3$ GeV,
and the supersymmetry breaking scale $\mu_1\sim 10^4$ GeV.  In this case, if
$B_{\mu_{\tau}}$ is chosen to be around electro-weak scale, $\mu_{\tau}$ will 
be very small, which can be achieved by choosing the coupling product
$\lambda_5\lambda_6\sim 10^{-4}$.  Phenomenologically it does not matter to 
have a small $\mu_{\tau}$.  In fact this is what we need as we will see in 
the following.  It should be noted that $L_{\tau'}$ and $H_d$ appear in the 
superpotential in different ways, so that the term $L_{\tau'}H_u$ cannot be 
rotated away.
\par
\vspace{1.0cm}
   It is necessary to discuss the scalar potential of the theory to see the 
sneutrino vevs.  In this model, besides field $Y$, the fields that can have 
non-vanishing vevs are the sneutrinos in the slepton doublets $A_i$ and the
neutral components of the Higgs doublets $\phi_u$ and $\phi_d$,
\begin{equation}
\langle A_i\rangle=\left(\begin{array}{c}
v_i\\0
\end{array} \right)~,~~~
\langle \phi_u\rangle=\left(\begin{array}{c}
0\\v_u
\end{array} \right)~,~~~
\langle \phi_d\rangle=\left(\begin{array}{c}
v_d\\0
\end{array} \right)~.
\end{equation} 
Sneutrino vevs are determined by the minimum of the following neutral 
potential,
\begin{equation}
V_n = V_n^H+2B_{\mu_{\tau}}\sum_iv_i v_u+M_A^2\sum_iv_i^2
+\frac{g_1^2+g_2^2}{4}\sum_i v_i^2(v_u^2+v_d^2)~,
\end{equation}
where $V_n^H$ has not been written explicitly which is the Higgs potential
irrelevant to sneutrinos.  The scalar lepton mass $M_A$ has been calculated in 
Ref. [14].  Neglecting the Yukawa contribution, 
$M_A^2=\displaystyle\frac{3}{8}(\frac{\alpha_2}{4\pi})^2\Lambda_S^2$ with 
$\Lambda_S^2=8\displaystyle\frac{\lambda_1^2\mu_1^4}{m_1^2}$.  $g_1$ and $g_2$ 
are the SU(2)$\times$U(1) coupling constants.  We expect $v_i\ll v_d$ or $v_u$ 
so as to keep the lepton universality.  Therefore in Eq. (11), all the terms 
of order $v_i^3$ and above have been dropped.  Straightforward analysis shows 
that 
\begin{equation}
v_1=v_2=v_3=-\frac{B_{\mu_{\tau}}v_u}{M_A^2+\frac{1}{2}M_Z^2\cos2\beta}~,
\end{equation}
where $\tan\beta=v_u/v_d$.  As we have seen that, even after the electro-weak
symmetry breaking, the $Z_3$ symmetry is still valid.  In other words, only 
$\tau'$-sneutrino has non-vanishing vev,
\begin{equation}
v_{\tau'}=-\frac{\sqrt{3}B_{\mu_{\tau}}v_u}{M_A^2
+\frac{1}{2}M_Z^2\cos2\beta}~,
~~~v_e=v_{\mu}=0~.
\end{equation}
Numerically $v_{\tau'}$ can be one order of magnitude lower than $v_d$, 
{\it e.g.} $v_{\tau'}\sim 10$ GeV by taking $M_A\sim 300$ GeV, 
$B_{\mu_{\tau}}\sim (50$ GeV)$^2$.  As has been mentioned before, 
non-vanishing $v_{\tau'}$ implies mixing between $\tau'$-neutrino and 
neutralinos.
\par
\vspace{1.0cm}
   The term $L_{\tau'}H_u$ provides a mixing mass between 
$(\nu_{\tau'}, \tau')$ and higgsinos.  The large $B_{\mu_{\tau}}$ can also 
cause comparatively large fermion mixing.  The $B_{\mu_{\tau}}$ term, which is
the mixing mass term between the slepton doublet $A_{\tau'}$ and Higgs doublet 
$\phi_u$, induces a renormalization to the corresponding fermion mixing mass 
term between $(\nu_{\tau'}, \tau')$ and the higgsino 
$(\tilde{\phi}_u^+, \tilde{\phi}_u^0)$ which is the superpartner of $\phi_u$.  
At one-loop level, this happens through Zino (for neutral fermion mixing) or 
Wino (for charged fermion mixing), $A_{\tau'}$ and $\phi_u$ being the virtual 
particles with $B_{\mu_{\tau}}$ insertion, as shown in Fig. 2.  The loop 
effect is approximately 
$\displaystyle\frac{g_2^2}{16\pi^2}\frac{B_{\mu_{\tau}}}{M_{\tilde{Z}}}$.  
Together with the contribution of $\mu_{\tau}$, the resulting fermion mixing 
mass $m_{\tau H}$ is 
\begin{equation}
m_{\tau H}\simeq \sqrt{3}\mu_{\tau}
+\frac{g_2^2}{16\pi^2}\frac{\sqrt{3}B_{\mu_{\tau}}}{M_{\tilde{Z}}}~.
\end{equation}
Requiring $B_{\mu_{\tau}}\sim (50$ GeV)$^2$ implies $\mu_{\tau}\sim 0.03$ GeV
due to Eq. (9).  Plus the loop effect, $m_{\tau H}\sim 0.04-0.1$ GeV.  
\par
\vspace{1.0cm}
   Let us now consider the neutral fermion mixing, namely the mixing of 
$\nu_{\tau}'$ and $\tilde{\phi}_u^0$.  This will give out the $\nu_{\tau}$ 
mass.  For this purpose, the full mass matrix of $\nu_{\tau'}$ and neutralinos
should be written down.  The Lagrangian for the neutralino masses is given as
\begin{equation}
-i(\nu_{\tau'}~~ \tilde{\phi}^0_d~~ \tilde{\phi}^0_u~~ \tilde{Z}~~ \tilde{X})
\left(\begin{array}{ccccc}
0          &0           &m_{\tau  H}   &av_{\tau'}   &0           \\
0          &0           &0             &av_d         &\lambda_4 v_u\\
m_{\tau H} &0           &0             &-av_u        &\lambda_4 v_d\\
av_{\tau'} &av_d        &-av_ u        &M_{\tilde{Z}}&0           \\
0          &\lambda_4v_u&\lambda_4v_d  &0            &0
\end{array}
\right)
\left(\begin{array}{c}
\nu_{\tau'}\\ \tilde{\phi}^0_d\\ \tilde{\phi}^0_u \\ \tilde{Z}\\ \tilde{X}
\end{array}
\right)
+{\rm h.c.},
\end{equation}
where $\tilde{\phi}_d$ and $\tilde{X}$ are the fermion components of $H_d$ and
$X$ respectively, $a=\displaystyle(\frac{g_1^2+g_2^2}{2})^{1/2}$.  The 
determinant of this matrix is approximately 
$2m_{\tau  H}a^2\lambda_4^2v_{\tau'}v_u(v_u^2+v_d^2)$ by taking 
$m_{\tau  H}\ll av_{\tau'}$.  Except for $\nu_{\tau}$, the masses of other 
neutralinos are at the electro-weak scale.  Therefore we have 
\begin{equation}
m_{\nu_{\tau}}\simeq \frac{m_{\tau  H}v_{\tau'}}{M_Z}~,
\end{equation}
which can be naturally within the range $(1-10)$ MeV.  The eigenstate is
\begin{equation}
\nu_{\tau}=N_{\nu}(\nu_{\tau'}-\frac{v_{\tau'}v_d}{v_d^2+v_u^2}
\tilde{\phi}^0_d+\frac{v_{\tau'}v_u}{v_d^2+v_u^2}
\tilde{\phi}^0_u)~,
\end{equation}
with $N_{\nu}$ being the normalization constant.
If the induced mass $m_{\tau H}$ were vanishing, it is easy to see that the
mass matrix in Eq. (15) would be of rank $4$ (instead of $5$), despite the
sneutrino vev is non-vanishing.
The absence of the conventional $\mu$ parameter is crucial for the
$\tau$-neutrino being very light compared to the weak scale.  If a weak
scale $\mu$ parameter were included, the neutrino mass would be at weak scale
or so.
\par
\vspace{1.0cm}
   The mixing of the $\tau'$ lepton with charginos is not so interesting as
that of neutralinos.  It just renormalizes slightly the $\tau$ lepton and 
chargino masses.  The related mass matrix is
\begin{equation}
(\tau^c~~ \tilde{\phi}_u^+~~ \tilde{W}^+)
\left(\begin{array}{ccc}
g_Yv_d       &-g_Yv_{\tau'} &0\\
m_{\tau  H}  &0             &g_2v_u\\
g_2v_{\tau'} &g_2v_d        &M_{\tilde{W}}\\
\end{array}
\right)
\left(\begin{array}{c}
\tau'^-\\ \tilde{\phi}_d^-\\ \tilde{W}^-
\end{array}
\right)~,\\[3mm]
\end{equation}
where $\tau^c$ is the charge conjugate field of the right-handed $\tau$ lepton
which has a Yukawa coupling $g_Y$ with $\tau'^-$.  At this stage, muon and
electron are still massless because of the $Z_3$ family symmetry.  The 
physical $\tau$ lepton state is 
\begin{equation}
\tau = N_{\tau} (\tau'-\frac{v_{\tau'}}{v_d}\tilde{\phi}_d^-)~,
\end{equation}
with $N_{\tau}$ being the normalization constant.
\par
\vspace{1.0cm}
   In this kind of supersymmetric model, $\tau$-neutrino only decays via 
$W$-boson exchange, $\nu_{\tau}\rightarrow e^+e^-\nu_e$.  The cosmology and 
astrophysics require a $\nu_{\tau}$ lifetime smaller than 100 sec [1].  That 
means a heavier $\nu_{\tau}$ is favored.  Taking $m_{\nu_{\tau}}=10$ MeV, the
lifetime is [19]
\begin{equation}
\begin{array}{lll}
\tau_{\nu_{\tau}}&\simeq& \displaystyle\frac{192\pi^3}{G_F^2m_{\nu_{\tau}}^5}
\frac{1}{|V_{e\tau}|^2}\\[3mm]
&\simeq& 0.3\displaystyle\times \frac{1}{|V_{e\tau}|^2}~{\rm sec}~.
\end{array}
\end{equation}
In this case, the $e-\tau$ CKM like mixing is required to be 
$|V_{e\tau}|\geq 0.05$ \footnote{This requirement is not totally unreasonable.  
Although the mass hirarchy of neutrinos is huge, that of charged leptons is 
not.  In certain extreme situation, the relation 
$V_{e\tau}\sim\sqrt{\frac{m_{\mu}}{m_{\tau}}}\sim 0.3$ might be hold.}.  This 
needs to be studied after including masses for $e$, $\mu$ and their neutrinos.
\par
\vspace{1.0cm}
   Phenomenologically, this model predicts lepton universality violation in 
the $\tau$ lepton decays.  It is because $\nu_{\tau}$ and $\tau$ in Eqs. (17) 
and (19) do not coincide in form.  Compared to the $e-\nu_e$ or 
$\mu-\nu_{\mu}$ weak transition, the $\tau-\nu_{\tau}$ transition amplitude is 
suppressed by a factor $N_{\nu}N_{\tau}(1+\frac{v_{\tau'}^2}{v_u^2+v_d^2})$.  
This factor can be effectively absorbed into the gauge interaction coupling 
constant $g^{\tau}$.  Therefore it just measures the $\tau$ lepton 
universality violation.  With reasonable choice of $\tan\beta$, like 
$\tan\beta\simeq 2.2$, for $v_{\tau'}\simeq 10$ GeV, the $e-\tau$ universality 
violation is at $10^{-3}$ level,
\begin{equation}
g^e:g^{\tau}=1:0.996~,
\end{equation}
which is still consistent with experiment [20], but near to the experiment 
limit.
\par
\vspace{1.0cm}
{\bf III. Summary and Discussions}\\

   In summary, we have described a supersymmetric model which can naturally 
accommodate MeV tau neutrino within the framework of GMSB.  The lepton number 
violation is introduced in the messenger sector of the theory, which then is 
communicated into the SM sector at one-loop level.  It turns out that a large
$B_{\mu_{\tau}}$ term and a small $\mu_{\tau}$ term (see Eqs. (7, 8)) are 
generated.  Furthermore, a non-vanishing sneutrino vev (see Eq. (13)) is 
produced.  These results cause, in an interesting manner, a non-zero 
$\nu_{\tau}$ mass which is right in the range of $(1-10)$ MeV.  Such a mass 
for tau neutrino and the phenomenological consequence for lepton universality
violation can be verified by the experiments in the near future.  
\par
\vspace{1.0cm}
   We have noted that this kind model is specific as far as the $\mu$-term is
concerned.  That term is not necessary in the model.  The electroweak 
symmetry is broken due to the introduction of the $\mu_2$ term which can be 
regarded as an expedient.  The explanation of $\mu_2$,
hence the radiative electroweak symmetry breaking, is beyond the scope of
this paper.  It is reasonable to discuss it when the $\mu$ problem in GMSB
gets a satisfactory understanding.
\par
\vspace{1.0cm}
   This model is of theoretical interest.  Firstly, the mechanism of
supersymmetry breaking is still an open problem.  MeV neutrino in GMSB is
worthy to be explored.  Secondly, although the MeV neutrino is not intrinsic
to the GMSB, the way to achieve it in this paper is very different from
that in the supergravity [11].  It allows naturally a rather large sneutrino
vev while the bilinear R-parity violation is small.  This scenario may have
other physical consequences, {\it e.g.} in the flavor problem [21].  A
detailed investigation on them is left for future works.
\par
\vspace{2.0cm}
\begin{center}
{\bf Acknowlegement}\\
\end{center}
   We would like to thank Manuel Drees for comments.  This work was supported 
in part by the Korea Science and Engineering Foundation (KOSEF) through the 
Center for Theoretical Physics at Seoul National University.

\newpage
\noindent
\begin{center} {\Large\bf FIGURE CAPTIONS} \end{center}
\vskip1cm
   Fig. 1
\hskip .3cm
Superfield diagrams for generating $\mu_{\tau}$ (a) and $B_{\mu_{\tau}}$ (b).
The internal lines with (without) a "$\times$" denote a messenger 
$\langle VV\rangle$ or$\langle SS\rangle$ ($\langle V^{\dagger}V\rangle$ or 
$\langle S^{\dagger}S\rangle$) propagator.  The field $Y$ can also be attached 
to the $V$ line.
\vskip0.5cm
   
   Fig. 2
\hskip .3cm
One-loop diagram for neutral fermion mixing due to the $B_{\mu_{\tau}}$ term
which is denoted as "$\times$".  $\tilde{Z}$ stands for Zino.

\newpage
{\Large\bf Figures}\\
\vskip 2cm
\begin{center}
\epsfig{file=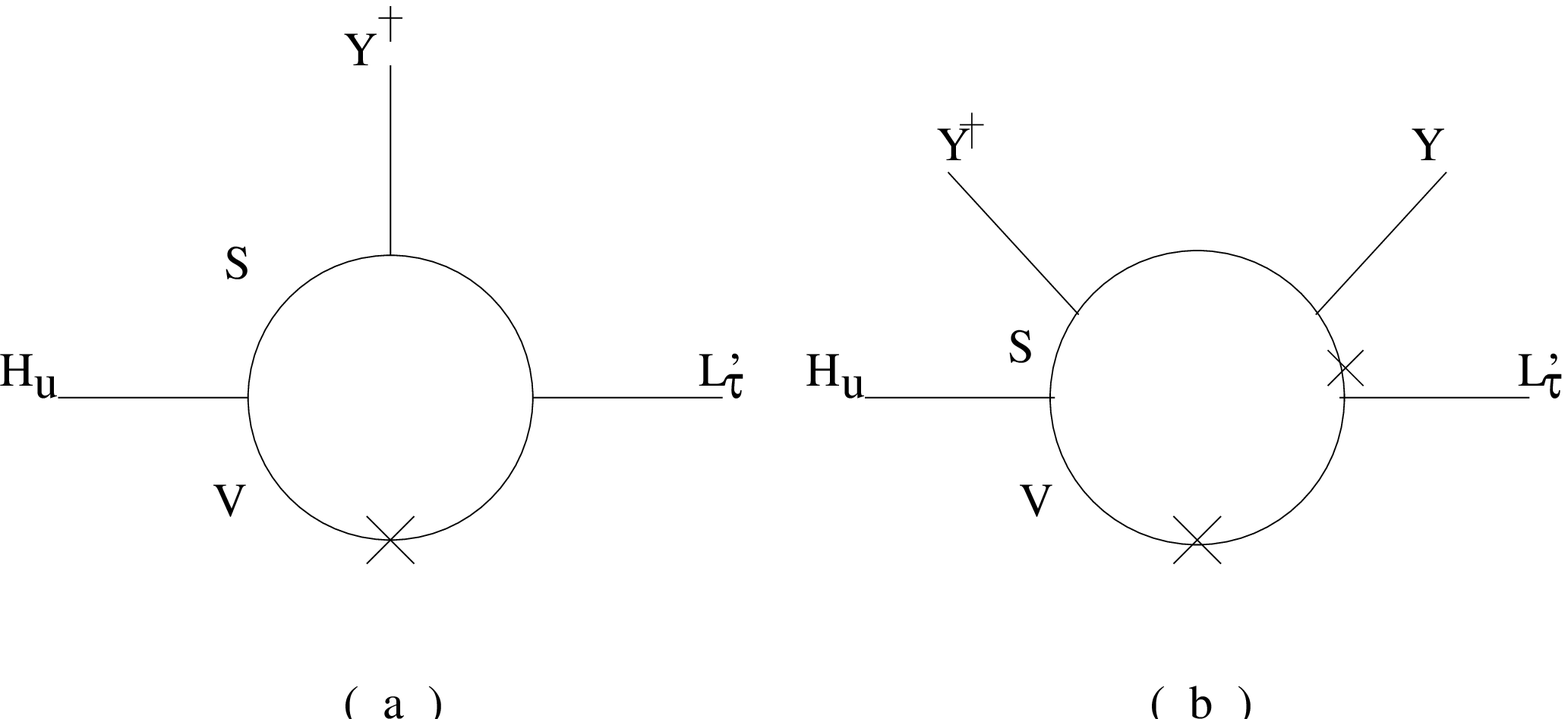, height=6cm}
\vskip.5cm
{\large\bf Fig.~1}
\end{center}

\newpage
\begin{center}
\vskip 2cm
\epsfig{file=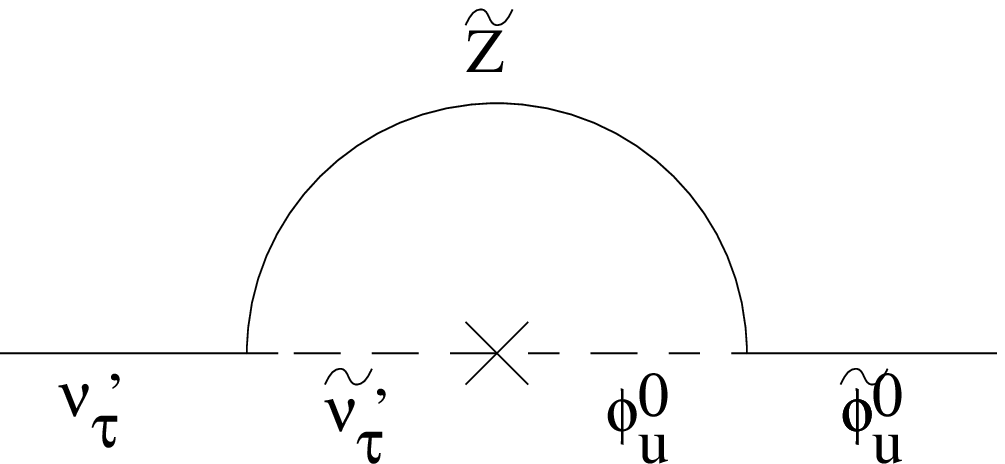,height=6cm}
\vskip.5cm
{\large\bf Fig.~2}
\end{center}


\newpage
\bigskip
\bigskip
\bigskip

\begin{thebibliography}{99}

\bibitem{}For a review, see G. Gyuk and M.S. Turner, Nucl. Phys. 
B(Proc. Suppl.) 38 (1995) 13.

\bibitem{}A.D. Dolgov, S. Pastor, J.C. Rom\~ao and J.W.F. Valle, Nucl. Phys.
B496 (1997) 24.

\bibitem{}K. Choi and A. Santamaria, Phys. Rev. D42 (1990) 293;\\
J.L. Feng, T. Moroi, H. Murayama and E. Schnapka, Phys. Rev. D57 (1998) 5875.\\

\bibitem{}C. Aulakh and R. Mohapatra, Phys. Lett. B119 (1982) 136;\\
G. Ross and J. Valle, Phys. Lett. B151 (1985) 375.

\bibitem{}T. Banks, Y. Grossman, E. Nardi and Y. Nir, Phys. Rev. D52 (1995) 
5319;\\
F. Vissani and A. Yu Smirnov, Nucl. Phys. B460 (1996) 37;\\
R. Hempfling, Nucl. Phys. B478 (1996) 3;\\
F.M. Borzumati {\it et al.}, Phys. Lett. B384 (1996) 123;\\
Y. Grossman and H.E. Haber, Phys. Rev. Lett. 78 (1997) 3438;\\
M. Hirsch, H.V. Klapdor-Kleingrothaus and S.G. Kovalenko, Phys. Lett. B398 
(1997) 311;\\
H.-P. Nilles and N. Polonsky, Nucl. Phys. B484 (1997) 33;\\
E. Nardi, Phys. Rev. D55 (1997) 5772;\\
S. Roy and B. Mukhopadhyaya, Phys. Rev. D55 (1997) 7020;\\
M. Bisset {\it et al.}, hep-ph/9804282.

\bibitem{}L. Hall and M. Suzuki, Nucl. Phys. 231 (1984) 419.

\bibitem{}S. Dimopoulos and L. Hall, Phys. Lett.  B207 (1988) 210;\\
K. Babu and R. Mohapatra, Phys. Rev. Lett. 64 (1990) 1705;\\
R. Barbieri, M. Guzzo, A. Masiero and D. Tommasini, Phys. Lett B252 (1990) 
251;\\
K. Enqvist, A. Masiero and A. Riotto, Nucl. Phys. B373 (1992) 95;\\
R.M. Godbole, P. Roy and X. Tata, Nucl. Phys. B401 (1993) 67.

\bibitem{}For a review, see G. Bhattacharyya, hep-ph/9709395.

\bibitem{}I.H. Lee, Phys. Lett. B138 (1984) 121, Nucl. Phys. B246 (1984) 
120;\\
B. de Carlos and P.L. White, Phys. Rev. D54 (1996) 3427.

\bibitem{}Y. Grossman, SLAC-PUB-7671, hep-ph/9710276.

\bibitem{}M.A. D\'iaz, J.C. Rom\~ao and J.W.F. Valle, Nucl. Phys. B524 (1998)
23;\\
For a review, see J.W.F. Valle, hep-ph/9808292.

\bibitem{}M. Dine and W. Fischler, Phys. Lett. B110 (1982) 227;\\
L. Alvarez-Gaum\'e, M. Claudson and M. Wise, Nucl. Phys. B207 (1982) 96;\\
C.R. Nappi and B.A. Ovrut, Phys. Lett. B113 (1982) 175.

\bibitem{}J. Ellis et al., Phys. Lett. B150 (1985) 142.

\bibitem{}M.C. Gonzalez-Garcia and J.W.F. Valle, Nucl. Phys. B355 (1991) 330.

\bibitem{}M. Dine and A.E. Nelson, Phys. Rev. D48 (1993) 1277;\\
M. Dine, A.E. Nelson and Y. Shirman, Phys. Rev. D51 (1995) 1362;\\
M. Dine, A.E. Nelson, Y. Nir and Y. Shirman, Phys. Rev. D53 (1996) 2658.

\bibitem{}G. Dvali, G.F. Giudice and A. Pomarol, Nucl. Phys. B478 (1996) 31.

\bibitem{}J.E. Kim and H.-P. Nilles, Phys. Lett. B138 (1984) 150.

\bibitem{}For reviews, see M. Dine, hep-ph/9707413;\\
C. Kolda, IASSNS-HEP-97/90, hep-ph/9707450.

\bibitem{}C.W. Kim and A. Pevsner, {\it Neutrinos in Physics and Astrophysics}
(Harwood Academic, 1993).

\bibitem{}W.J. Marciano, Nucl. Phys. B(Proc. Suppl.) 40 (1995) 587;\\
A. Pich, Nucl. Phys. B(Proc. Suppl.) 55C (1997) 3;\\
K. Hagiwara, KEK-TH-461 (Dec, 1995).

\bibitem{}D. Du and C. Liu, Mod. Phys. Lett. A8 (1993) 2271; A10 (1995)
1837;\\
C. Liu, Int. J. Mod. Phys. A11 (1996) 4307; Mod. Phys. Lett. A12 (1997) 329.

\end{thebibliography}
\end{document}